\documentclass[aps,prd,tightenlines,showpacs,amsmath,amssymb]{revtex4}%
\usepackage{graphicx}
\usepackage{epsfig}
\usepackage{graphics,color}
\usepackage{amssymb}
\usepackage{hyperref}
\usepackage{amsbsy}
\usepackage{bm}
\newcommand{\be}{\begin{equation}}
\newcommand{\ee}{\end{equation}}
\newcommand{\bse}{\begin{subequations}}
\newcommand{\ese}{\end{subequations}}
\newcommand{\bea}{\begin{eqnarray}}
\newcommand{\eea}{\end{eqnarray}}

\newcommand{\bean}{\begin{eqnarray*}}
\newcommand{\eean}{\end{eqnarray*}}

\begin{document}
\title{Chiral soliton model at finite temperature and density}
\author{Hong Mao$^{1,2}$}
\email {mao@hznu.edu.cn}
\author{Tianzhen Wei$^{3}$}
\author{Jinshuang Jin$^{1}$}
\address{1.Department of Physics, Hangzhou Normal University, Hangzhou 310036, China \\
2.Theoretical Research Division, Nishina Center, RIKEN, Saitama 351-0198, Japan\\
3.Science Division, Hangzhou NO.15 Middle School, Hangzhou 310013, China}


\begin{abstract}
In mean field approximation, we study a chiral soliton of the linear sigma model with two flavors at finite temperature and density. The stable soliton solutions are calculated with some appropriate boundary conditions. Energy and radius of the soliton are determined in a hot medium of constituent quarks. It is found that for $T<T_c$, the energy of the soliton $E^*$ is less than the energy of three free constituent quarks $3M_q$, but with the increasing of temperature, the difference between $E^*$ and $3M_q$ becomes smaller and smaller, once $T>T_c$, there is a sharp delocalization phase transition from hadron matter to quark matter coincident with the restoration of chiral symmetry. In the transition region, the thermodynamic properties show large discontinuities which is an indication for a first-order phase transition.
\end{abstract}


\maketitle

\section{Introduction}
It is widely believed that at sufficiently high temperatures and densities there is a quantum chromodynamics (QCD) phase transition between normal nuclear matter and quark-gluon plasma (QGP), where quarks and gluons are no longer confined in hadrons\cite{Rischke:2003mt}\cite{Yagi:2005yb}. Experimentally, the study of the QCD phase transition is supported by the heavy-ion collisions in laboratories at ultrarelativistic energies, such as the Relativistic Heavy-Ion Collider (RHIC) at Brookhaven National Laboratory and the Large Hadron Collider (LHC) at CERN. These conducted experiments provide us with a chance to create a hot QCD matter and elucidate its properties. In order to explore of a wider range of the QCD phase transition up to several times of the normal nuclear matter density, the new Facility for Antiproton and Ion Research (FAIR) at Darmstadt, the Nuclotron-based Ion Collider Facility (NICA) at the Joint Institute for Nuclear Research (JINR) in Dubna and the Japan Proton Accelerator Research Complex (J-PARC) at JAEA and KEK, will make such extreme conditions possible through collisions\cite{Fukushima:2010bq}.

On the theoretical side, due to the property of confinement, in nonperturbative regime of large distances, or equivalently low energies, the analytical as well as numerical methods have not been developed enough to fulfill the solution of low-energy nonperturbative cases, especially if the baryons are involved. Therefore the challenge to nuclear physicists is to find models which can bridge the gap between the fundamental theory and our wealth of knowledge about low energy phenomenology, and these models should be successful in explaining empirical facts at low energies, for example, the dynamical breaking of chiral symmetry and the confinement,which are both intimately related to the nonperturbative structure of the QCD vacuum.

The linear sigma model (LSM)\cite{GellMann:1960np} for the phenomenology of QCD has been proposed to describe the vacuum structure with incorporating chiral symmetry and its spontaneous breaking. This model has two charming features according to two distinct low-energy phenomena. On the one hand, if we assume the thermal medium is mimicked by a uniform constituent quark medium with a dynamically generated mass, the model can be used to describe a restoration of the chiral symmetry at finite temperature and density in satisfactory agreement with the lattice calculations\cite{DeTar:2009ef}\cite{Philipsen:2012nu} and of the Nambu-Jona-Lasinio (NJL) model as well\cite{Vogl:1991qt}. On the other hand, starting from the same Lagrangian, bound states (chiral solitons) of valence quarks can be constructed through the interaction with $\sigma$ and $\pi$ mesons\cite{Birse:1983gm,Kahana:1984dx}, and the nucleon naturally arises as a non-topological chiral soliton in the model. Moreover, the model has proven to be a successful approach to the description of nucleon static properties in vacuum\cite{Birse:1983gm,Alberto:1988xj,Bernard:1988db,Alberto:1990ru,Goeke:1988hp,Aly:1998wg}. Combining these two features together by requiring a soliton embedded in the hot soup of constituent quarks, the model seems to provide a suitable working scheme to simultaneously study both the restoration of chiral symmetry and the possible dissolution of the soliton, which simulates the delocalization (or deconfinement in some literatures) transition of nuclear matter to quark matter. Thus such a chiral soliton model is obviously more advantageous than other nontopological soliton models\cite{Friedberg:1976eg}\cite{Wu:2005ty}\cite{Birse:1991cx} in the description of the hadron-quark phase transition\cite{Mao:2006zp} \cite{Reinhardt:1985nq}.

Using this model, the nucleon as a $B=1$ chiral soliton in a cold quark medium has been investigated in Ref\cite{Christov:1991ry}. In their studies, the nucleon is treated as a chiral soliton of the LSM, whereas the parameters $f_{\pi}$, $m_{\pi}$ and $m_{\sigma}$ are chosen to be the medium-modified meson values within the NJL model. For finite temperature, more recently, by adopting the one-loop phenomenological mesonic potential\cite{Hong1997clj} and the coherent-pair approximation\cite{Goeke:1988hp}\cite{Aly:1998wg}, Abu-Shady and Mansour have studied nucleon properties at finite temperature\cite{AbuShady:2012zza}. Since the equally one-loop contribution of the quarks in potential is ignored, their studies can not be applied to finite density directly. In order to get a full picture of hadron-quark phase transition either at finite temperature or finite density, in present work, in mean field approximation, we are going to investigate the nucleon properties as well as the thermodynamics of the system at finite temperature and density. As we know the NJL model and the LSM  treat the contribution of the Dirac sea differently, in the NJL model it is included explicitly up to a momentum cutoff $\Lambda$, while in the LSM this contribution is renormalized out. So that Different from the chiral soliton in the NJL model\cite{Berger:1996hc}\cite{Schleif:1997pi}, the present model has the benefit of renormalizability. Eventually, such a model would be investigated beyond mean field by the loop expansion, especially by using the CJT effective potential\cite{Rischke:2003mt}\cite{Cornwall:1974vz}.

The structure of the paper is as follows: in the next section we introduce the chiral soliton model with two quark flavors, nucleon static properties in the vacuum are briefly discussed and parameters are fixed. Then in mean field approximation, the chiral soliton solutions of the model at different temperatures and densities are self-consistently solved in Sect. III. Section IV is devoted to study static properties of nucleon at finite temperature and density, but we leave the study of the hadron-quark phase transition until section V. At the end, we give discussions and summary in Sec.VI.

\section{The Model}
The chiral effective Lagrangian of the $SU(2)_{R}\times SU(2)_{L}$ symmetry linear sigma model with two quark flavors has the form\cite{GellMann:1960np}\cite{Scavenius:2000qd}
\begin{equation}
{\cal L}=\overline{\psi} \left[ i\gamma ^{\mu}\partial _{\mu}-
g(\sigma +i\gamma _{5}\vec{\tau} \cdot \vec{\pi} )\right] \psi
+ \frac{1}{2} \left(\partial _{\mu}\sigma \partial ^{\mu}\sigma +
\partial _{\mu}\vec{\pi} \cdot \partial ^{\mu}\vec{\pi}\right)
-U(\sigma ,\vec{\pi}) \,\, ,
\label{Lagrangian}
\end{equation}
here we have introduced a flavor-blind Yukawa coupling $g$
of the isodoublet, spin-$\frac{1}{2}$ quark fields $\psi=(u,d)$ to interact with the spin-$0$,
isosinglet $\sigma$ and the isotriplet pion field $\vec{\pi} =(\pi _{1},\pi _{2},\pi _{3})$. The potential for the $\sigma$ and $\vec{\pi}$
is parametrized as
\begin{equation}
U(\sigma ,\vec{\pi})=\frac{\lambda}{4} \left(\sigma ^{2}+\vec{\pi} ^{2}
-{\vartheta}^{2}\right)^{2}-H\sigma -\frac{m^{4}_{\pi}}{4 \lambda}+f^{2}_{\pi}m^{2}_{\pi},
\label{mpot}
\end{equation}
and the minimum energy occurs for chiral fields $\sigma$ and $\vec{\pi}$ restricted to the chiral circle in the physical vacuum:
\begin{eqnarray}
\sigma^2+\vec{\pi}^2=f_{\pi}^2,
\label{ccircle}
\end{eqnarray}
where $f_{\pi}=93 \mathrm{MeV}$ is the pion decay constant and $m_{\pi}=138$ MeV is the pion mass, the last two constant terms in Eq.(\ref{mpot}) are used to guarantee that the energy of vacuum in the absence of quarks is zero. The constant $H$ is fixed by the PCAC relation which gives $H=f_{\pi}m_{\pi}^{2}$. We further assume that the chiral symmetry is spontaneously broken in the vacuum and the expectation values of the meson fields are $\langle\sigma\rangle ={\it f}_{\pi}$ and $\langle\vec{\pi}\rangle =0$, then the dimensionless coupling constants $g$ and $\lambda$ are the only two free parameters of the model, which are in turn conveniently re-expressed in terms of the constituent quark mass in vacuum, $M_{q}=gf_\pi$, and the sigma mass, $m_\sigma^2=m^2_\pi+2\lambda f^2_\pi$. Finally, the quantity $\vartheta^2$ can subsequently be expressed as $\vartheta^{2}=f^{2}_{\pi}-m^{2}_{\pi}/\lambda$.

The mass of the sigma meson is still a poorly known number, but the most recent compilation of the Particle Data Group considers that $m_{\sigma}$ can vary from $400$ MeV to $550$ MeV with full width $400-700$ MeV\cite{Beringer:1900zz}. The coupling constant $g$ is usually fixed by the constituent quark mass in vacuum within the range of $300\sim 500$ MeV, which gives $g \simeq 3.3\sim 5.3$. In this work we take $m_{\sigma}=472$ MeV and $g=4.5$ as the typical values in order to describe the properties of nucleon in vacuum successfully.

In vacuum, the $\sigma$ and $\pi$ are taken as time-independent, classical $c$-number fields, which only differ from their vacuum values in the neighborhood of the quark sources. The state of the quarks $\{\phi_n(\mathbf{r}) \}$ with energy $\{\epsilon_n\}$ and the $\sigma(\mathbf{r})$, $\pi(\mathbf{r})$ meson fields satisfy the coupled set of the Euler-Lagrange equations of motion
\begin{eqnarray}
-i \vec{\alpha}\cdot \vec{\nabla}\phi_n(\mathbf{r})-g \beta \left[\sigma(\mathbf{r})+i \gamma_5 \vec{\tau}\cdot \vec{\pi}(\mathbf{r}) \right] \phi_n(\mathbf{r})=\epsilon_n \phi_n(\mathbf{r}), \label{eom1}\\
-\nabla^2 \sigma (\mathbf{r})+\frac{\partial U(\sigma ,\vec{\pi} )}{\partial \sigma}=-g \sum_{n_{occ}} \bar{\phi}_n(\mathbf{r}) \phi_n(\mathbf{r}) \label{eom2}\\
-\nabla^2 \vec{\pi} (\mathbf{r})+\frac{\partial U(\sigma ,\vec{\pi} )}{\partial \vec{\pi}}=-g \sum_{n_{occ}} \bar{\phi}_n(\mathbf{r})i \gamma_5 \vec{\tau} \phi_n(\mathbf{r})
\label{eom3}
\end{eqnarray}
with
\begin{eqnarray}
\int \phi^{\dag}_n(\mathbf{r}) \phi_n(\mathbf{r}) d^3 r=1
\end{eqnarray}
where $\vec{\alpha}$ and $\beta$ are the conventional Dirac matrices.

The ground state of the chiral soliton is the state where $N$ quarks in the same lowest Dirac state $\phi_0$ with energy $\epsilon$. In the following, our discussions are constrained in the case of $N=3$ for baryons. In order to obtain solutions of minimum energy, we adopt the ``hedgehog" ansatz with the meson fields are spherically symmetric and valence quarks are in the lowest s-wave level
\begin{eqnarray}
\sigma = \sigma(r), \vec{\pi} = \hat{\mathbf{r}}\pi(r),\\
\phi_0 = \left(\begin{array}{c} u(r) \\
i \vec{\sigma} \cdot\mathbf{\hat{ r}}v(r)
\end{array}\right)\chi, \label{sltconf}
\end{eqnarray}
where $\chi$ is a state in which the spin and isospin of the quark couple to zero:
\begin{eqnarray}
(\vec{\sigma}+\vec{\tau})\chi=0.
\end{eqnarray}
Now the system is spherical symmetric and the Euler-Lagrange equations of motion (\ref{eom1})-(\ref{eom3}) transform in radial coordinates to
\begin{eqnarray}
\frac{du(r)}{dr} =  -\left(\epsilon+g \sigma(r)\right)v(r)-g \pi(r)u(r), \label{equation1}\\
\frac{dv(r)}{dr} =  -\left(\frac{2}{r}-g\pi(r)\right)v(r)+\left(\epsilon-g \sigma(r)\right)u(r), \label{equation2} \\
\frac{d^2 \sigma(r)}{dr^2}+\frac{2}{r}\frac{d\sigma(r)}{dr}-\frac{\partial U}{\partial \sigma}= Ng\left(u^2(r)-v^2(r)\right),\label{equation3}\\
\frac{d^2 \pi(r)}{dr^2}+\frac{2}{r}\frac{d\pi(r)}{dr}-\frac{2 \pi(r)}{r^2}-\frac{\partial U}{\partial \pi} = -2Ng u(r) v(r),
\label{equation4}
\end{eqnarray}
and the quark functions should satisfy the normalization condition
\begin{eqnarray}\label{norm}
4\pi \int r^2 \left(u^2(r)+v^2(r)\right)dr=1.
\end{eqnarray}
These equations are subject to the boundary conditions which follow
from the requirement of finite energy:
\begin{eqnarray}
v(0)=0,     \frac{d\sigma(0)}{dr}=0,\pi(0)=0 \label{boundary1}\\
u(\infty)=0,\sigma(\infty)={\it f}_{\pi},\pi(\infty)=0.\label{boundary2}
\end{eqnarray}
The asymptotic vacuum value of the soliton field has to be determined by an additional condition that the physical vacuum is recovered at infinity. In this ``physical" vacuum the quarks are free Dirac particles of the constituent mass $g \sigma_v$, and the chiral symmetry is spontaneously broken.

If we put $N$ quarks into the lowest state with energy $\epsilon$, the total energy of the hedgehog baryon is given by
\begin{eqnarray}\label{energy}
E=N \epsilon+4\pi\int r^2 \left[ \frac{1}{2}
\left(\frac{d\sigma}{dr}\right)^2+\frac{1}{2}\left(\frac{d\pi}{dr}\right)^2+U(\sigma ,\pi) \right]dr.
\end{eqnarray}

The model has two adjustable parameters $g$ and $\lambda$ which can be chosen to fit various baryon properties, such as masses, charge radii and magnetic moments. Once the solutions to the above equations are obtained, one can calculate these physical quantities pertaining to the three-quark system, which have been measured experimentally.

\begin{figure}
\includegraphics[scale=0.36]{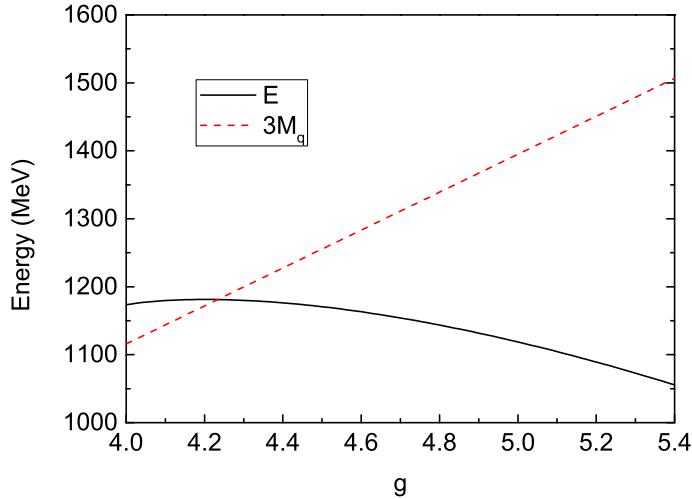}
\caption{\label{Fig01} (Color online) The total energy of system $E$ and the energy of $3$ free constituent quark $3M_q$ as a function of the Yukawa coupling $g$. For $g<g_c$, $E$ is lager than $3 M_q$, there is no stable bound state of quarks, however, for $g\geq g_c$, the soliton solution is stable, where $g_c=4.24$.}
\end{figure}

Unlike the Friedberg-Lee model\cite{Friedberg:1976eg} and its descendant models\cite{Birse:1991cx}, where the confinement of quarks is approximated through their interaction with the phenomenological scalar field $\sigma$ which is introduced to describe the complicated nonperturbative features of the QCD vacuum. In many of these models, at large radius $r$, the $\sigma$ field assumes its vacuum value $\sigma_v$, but at small $r$, the $\sigma$ field has a value close to the second minimum of the potential near zero. It means that in physical vacuum state, the quark mass is more than $1 \mathrm{GeV}$ which makes it energetically unfavorable for the quark to exist freely, so that the effective heavy quarks have to be confined in hadron bags. Similar to the MIT bag model, they need to introduce the bag constant to make the baryon state stable. However, for the chiral soliton model, the interaction between the meson fields and the quarks is essential for the formation of a stable soliton, the state could be bound only when the total energy of system is lower than the energy of three free constituent quarks in system. Hence the coupling constant $g$ or $M_q$ is the significant parameter in the present model. In Fig.\ref{Fig01} It is numerically shown that the critical value of coupling constant for $m_{\sigma}=472$ MeV is $g_c= 4.24$, below that there is no stable bound state of quarks.

\begin{figure}
\includegraphics[scale=0.36]{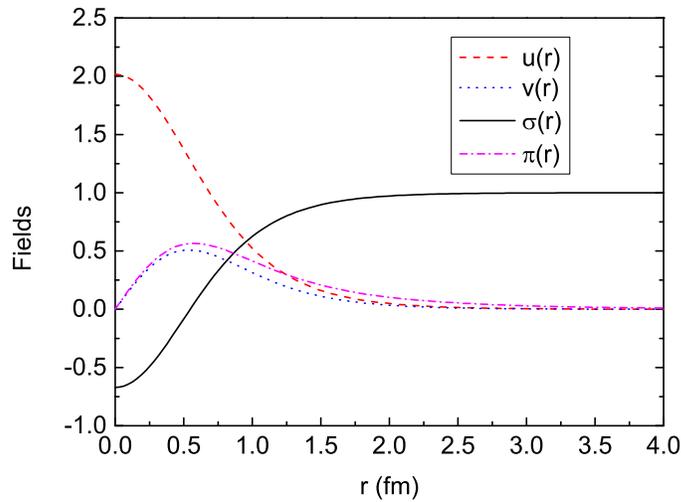}
\caption{\label{Fig02} (Color online) The quark fields in relative unit and the $\sigma$, $\pi$ fields scaled with $f_{\pi}$ as function of the radius $r$ for the parameters taken as $m_{\sigma}=472$ MeV and $g=4.5$.}
\end{figure}

Here after taking the set of parameters $m_{\sigma}=472$ MeV and $g=4.5$, we plot the $\sigma$, $\pi$ and quark fields profiles in arbitrary unit as functions of radius $r$ in Fig.\ref{Fig02} and calculate the properties of the nucleon, such as $E=1170$ MeV, $R=0.877$ fm, $\mu_p=0.319 e$fm and $g_A/g_V=1.24$. For comparing we adopt experimental values for the proton as follows\cite{Beringer:1900zz}: $E=(M_p+M_{\Delta})/2=1085$ MeV, $R_p=0.877$ fm, $\mu_p=0.294 e$fm and $g_A/g_V=1.25$. Therefore it is proven that this set of parameters can describe the properties of nucleon at zero temperature in a reasonable way.

\section{Mean field Approximation}
A convenient framework of studying phase transitions is the thermal field theory. Within this framework, the finite temperature effective potential is an important and useful theoretical tool. In this section, in order to investigate the temperature and the chemical potential dependence of the chiral soliton, let us consider a spatially uniform system in thermodynamical equilibrium at temperature $T$ and quark chemical potential $\mu$. In general, the grand partition function reads
\begin{eqnarray}
\mathcal{Z}&=& \mathrm{Tr exp}[-(\hat{\mathcal{H}}-
\mu \hat{\mathcal{N}})/T] \nonumber \\
&=& \int\prod_a \mathcal{D} \sigma \mathcal{D} \pi_a \int
\mathcal{D}\psi \mathcal{D} \bar{\psi} \mathrm{exp} \left[ \int_x
(\mathcal{L}+\mu \bar{\psi} \gamma^0 \psi )
\right],
\end{eqnarray}
where $\int_x\equiv i \int^{1/T}_0 dt \int_V d^3x$, $V$ is the volume of the system and $\mu=\mu_B /3 $ for the homogeneous background field.

We evaluate the partition function in the mean-field approximation similar to the work of \cite{Scavenius:2000qd}. Thus we replace the meson fields by their expectation values in the action. In other words, we neglect both quantum and thermal fluctuations of the meson fields. The quarks and antiquarks are retained as quantum fields. The integration over the fermions yields a determinant which can be calculated by standard methods\cite{Kapusta:2006pm}. This generates an effective potential for the mesons. Finally, we obtain the thermodynamical potential density as
\begin{eqnarray}\label{potential}
\Omega(T,\mu)=\frac{-T \mathrm{ln}
\mathcal{Z}}{V}=U(\sigma ,\vec{\pi} )+\Omega_{\bar{\psi}
\psi},
\end{eqnarray}
with the quarks and antiquarks contribution
\begin{eqnarray}
\Omega_{\bar{\psi} \psi} &=& -\nu_q T \int \frac{d^3\vec{p}}{(2
\pi)^3} \left\{ \mathrm{ln} \left[ 1+e^{-(E_q-\mu)/T}\right] +\mathrm{ln} \left[
1+e^{-(E_q+\mu)/T}\right]\right\}.
\end{eqnarray}
Here, $\nu_q=2N_f N_c=12$ and $E_q=\sqrt{\vec{p}^2+M_q^2}$ is the valence quark and antiquark energy for $u$ and $d$ quarks, and the minus sign is the consequence of Fermi-Dirac statistics. The constituent quark (antiquark) mass $M_q$ is defined as
\begin{equation}
M^2_q=g^2\left(\sigma^2+\vec{\pi}^2 \right)=g^2 \sigma_v^2,
\label{qmass}
\end{equation}
where, $\sigma_v \equiv \sqrt{\sigma^2+\vec{\pi}^2}$ is a temperature and density-dependent chiral parameter, which is introduced to characterize the chiral symmetry breaking during the QCD phase transition. Traditionally, $\sigma$ and $\vec{\pi}$ fields in above equation are independent of space and time. In the physical vacuum, the expectation value of the pion field is set to zero, $\vec{\pi}=0$, and thus $M^2_q=g^2 \sigma^2=g^2 \sigma_v^2$. Then the value of $\sigma_v$ and thereby the quark mass in thermal background are determined by minimizing the thermodynamical potential in Eq.(\ref{potential}) with respective to these spacetime-independent meson fields $\sigma$ and $\pi$ in the physical vacuum. Through such a standard procedure, we can generate a uniform constituent quark thermal medium with a universal dynamical-generated mass $M_q$ where the soliton is going to be inserted. 

On the chiral soliton background, however, the above well-developed formula in mean field approximation can not be directly applied yet. Since in the calculation of this effective potential, the plane-wave valence quark states have been used, whereas, in fact, the bounded quarks should be confined in the finite-size solitonic configuration. In consequence, the thermodynamical potential obtained in Eq.(\ref{potential}) is independent of the space and does not contain any information about the soliton solution. Thus, the thermodynamical potential should be appropriately modified beforehand in order to allow the effect of thermal background with temperature $T$ and density $\mu$ to be included in the set of equations of motion under some proper requirements.    

As mentioned in above discussion, as long as the coupling constant $g$ is large enough, the minimum energy in a vacuum occurs for chiral fields restricted to the chiral circle in the physical vacuum. Such a constraint can be generalized to a more general case in the presence of the temperature and density by just replacing the $f_{\pi}$ in Eq.(\ref{ccircle}) with a chiral parameter $\sigma_v$ in the physical vacuum. By defining $\varphi(r) \equiv \sqrt{\sigma^2(r)+\vec{\pi}^2(r)}$, the general constraint has 
\begin{eqnarray}
\varphi(r)= \sqrt{\sigma^2(r)+\vec{\pi}^2(r)} \rightarrow \sigma_v \qquad \mathrm{for} \quad r\rightarrow \infty.
\label{chiralcircle}
\end{eqnarray}
From this new constraint, the constraint in vacuum can be naturally recovered as $\sigma_v=f_{\pi}$ for $T=\mu=0$. Moreover, similar to the results shown in Fig.(\ref{Fig02}), the $\varphi(r)$ can be taken as a constant $\sigma_v$ for large $r$. Only when $r$ is around the soliton radius $R$, the $\varphi(r)$ starts to slightly deviate from this constant. For a good approximation, we can assume the valence quark (antiquark) almost possesses with the same constituent mass in the whole space. This makes our discussions more reasonable than previous studies in the Friedberg-Lee model in vacuum or in thermal medium\cite{Reinhardt:1985nq}, where the free quark carries the current mass inside the soliton bag, whereas outside the soliton bag, the confined quark has an unphysical large masses. 

Equipped with the new constraint in Eq.(\ref{chiralcircle}), the modified thermodynamical potential $\Omega'(T,\mu)$ can be constructed by assuming a physical picture in which the valence quarks bearing a hedgehog configuration in the presence of the temperature and density, is simulated by these valence quarks facing the modified meson fields in thermal medium\cite{Christov:1991ry}. Thus, the $\Omega'(T,\mu)$ can be characterized as
\begin{eqnarray}\label{npotential}
\Omega'(T,\mu)=U(\sigma ,\vec{\pi})-\nu_q T \int \frac{d^3\vec{p}}{(2
\pi)^3} \left\{ \mathrm{ln} \left[ 1+e^{-(E'_q-\mu)/T}\right] +\mathrm{ln} \left[
1+e^{-(E'_q+\mu)/T}\right]\right\},
\end{eqnarray}
with $E'_q=\sqrt{\vec{p}^2+{M'}_q^2}$ and ${M'}^2_q=g^2 \varphi^2(r)=g^2\left(\sigma^2(r)+\vec{\pi}^2(r) \right)$. Furthermore, if we approximately treat a hot and dense thermal medium  as a uniform constituent quark medium with solitons embedded in, a new set of coupled equations of motion for the chiral soliton could be derived by simply replacing the relevant classical potential $U(\sigma ,\vec{\pi})$ with the appropriately modified thermal effective potential $\Omega'(T,\mu)$. Accordingly, a set of coupled equations for mesons can be described as\cite{Mao:2006zp}\cite{Scavenius:2000bb}\cite{Carter:2002te}

\begin{eqnarray}
\frac{d^2 \sigma(r)}{dr^2}+\frac{2}{r}\frac{d\sigma(r)}{dr}-\frac{\partial\Omega'}{\partial\sigma} = Ng\left(u^2(r)-v^2(r)\right),\label{equation5}\\
\frac{d^2 \pi(r)}{dr^2}+\frac{2}{r}\frac{d\pi(r)}{dr}-\frac{2 \pi(r)}{r^2}-\frac{\partial\Omega'}{\partial\pi} = -2Ng u(r) v(r),
\label{equation6}
\end{eqnarray}
where
\begin{eqnarray}
\frac{\partial \Omega' }{\partial \sigma}&=& \lambda\left(\sigma^2(r)+\pi^2(r)-\vartheta^2\right)
\sigma(r)-H+g\rho_{s}(r), \\
\frac{\partial \Omega'}{\partial \pi} &=& \lambda\left(\sigma^2(r)+\pi^2(r)-\vartheta^2\right)
\pi(r)+g \rho_{ps}(r).
\label{smmass}
\end{eqnarray}
The scalar and pseudoscalar densities of valence quarks and antiquarks can be expressed as
\begin{eqnarray}
\rho_s(r)= g \sigma(r) \nu_q \int \frac{d^3{\vec{p}}}{(2\pi)^3}
\frac{1}{E'_q}\left[\frac{1}{1+e^{(E'_q-\mu)/T}}+\frac{1}{1+e^{(E'_q+\mu)/T}}\right],\label{pscaldens}\\
\rho_{ps}(r)=g \pi(r) \nu_q \int \frac{d^3{ \vec{p}}}{(2\pi)^3}
\frac{1}{E'_q}\left[\frac{1}{1+e^{(E'_q-\mu)/T}}+\frac{1}{1+e^{(E'_q+\mu)/T}}\right].
\label{scaldens}
\end{eqnarray}
From the equations (\ref{equation5}) and (\ref{equation6}), for large $r$, if we require the $\sigma(r)$ asymptotically approaches to the expectation value $\sigma_v$ in the physical vacuum while setting other fields to zero, we can rediscover the well-known gap-equations presented in Ref.\cite{Scavenius:2000qd}, which should stand for the boundary conditions in the physical vacuum. On the contrary, for a small $r$, now that the chiral fields need not necessarily to be restricted to the chiral circle as the situation in physical vacuum, the bounded constituent quarks could develop various dynamical-generated masses at different radii. By combining the equations of motion for the quarks (\ref{equation1}) and (\ref{equation2}), we will finally obtain a set of coupled Euler-Lagrange equations of motion in the presence of the temperature and density for the chiral soliton. Moreover, these equations are automatically satisfied the requirement that the physical vacuum state must be realized when $r$ is infinite, so does for the vacuum state.     

Now, for $\mu=0$, the set of equations (\ref{equation1}),(\ref{equation2}),(\ref{equation5}) and (\ref{equation6}) with the normalization condition equation (\ref{norm}) can be solved for the soliton fields as functions of radius at finite temperature $T$ under some appropriate boundary conditions, and the solutions of these coupled equations determine the behavior of the chiral order parameter $\sigma_v$ as a function of $T$. Therefore, the properties of a soliton embedded in a thermal medium can be investigated with given temperature $T$. 

However, the situation becomes more complicated when we consider the soliton embedded in a thermal medium with finite chemical potential $\mu$. In such case, besides the bounded constituent quarks bearing the solitonic configurations, the unbound constituent quarks treated as the homogeneous background thermal fields with $T$ and $\mu$ will bring an additional contribution to the total baryon density as long as they are allowed to penetrate into the soliton by the requirement of the equations of motion of the soliton. Since these constituent quarks are unbouned and have a uniform constituent quark masses $M_q$, by using the standard thermodynamic relationship, the net baryon density of this additional contribution is given by
\begin{eqnarray}
\rho^m_B= -\frac{1}{3}\frac{\partial \Omega}{\partial \mu}= \frac{\nu_q}{3} \int \frac{d^3{\vec{p}}}{(2\pi)^3}
\left[\frac{1}{1+e^{(E_q-\mu)/T}}-\frac{1}{1+e^{(E_q+\mu)/T}}\right],
\end{eqnarray}
here, the $\rho^m_B$ also represents a homogeneous medium density respected to the unique chemical potential $\mu$. As a result, the soliton baryon density is split into valence and medium parts
\begin{eqnarray}\label{density}
\rho_B= \sum_{n_{occ}} \phi_n(\mathbf{r})^{\dag}\phi_n(\mathbf{r})+\rho^m_B.
\end{eqnarray}

In order to ensure the solitonic baryon number exactly to one, the normalization condition equation (\ref{norm}) should be modified as
\begin{eqnarray}\label{norm2}
4\pi \int r^2 \left(u^2(r)+v^2(r)\right)dr=1-B_m,
\end{eqnarray}  
with $B_m=4\pi \int_V \rho^m_B r^2 dr$ and $V$ bing the volume of the soliton. At last, the properties of a soliton embedded in a thermal medium can be investigated in both cases of zero and finite chemical potential by solving the set of equations (\ref{equation1}),(\ref{equation2}),(\ref{equation5}) and (\ref{equation6}). Moreover, from the equation (\ref{norm2}) we can naturally recover the result in vacuum or $\mu=0$ but finite $T$. This method is somewhat similar to the method presented in Ref\cite{Berger:1996hc} by introducing a different chemical potential for the soliton, since the effects of the first term in Eq.(\ref{density}) can be totally absorbed to the second term by redefining the chemical potential $\mu_s=\mu+\delta \mu$. But this leads to some difficulties if the different chemical potential $\mu_s$ is included. For example, if we take $\mu_s$ as a homogeneous chemical potential in space, a finite fraction of the baryon number is homogeneously spread outside the soliton, such a soliton is spatially unlimited. On the contrary, if we take $\mu_s$ as a space-dependent chemical potential, the standard way to derive the thermodynamical potential presented above is going to break down. So that, in this work, we prefer to choose the scheme by fixing the baryon number of the soliton to one through the modification of the normalization condition equation.   

\section{Nucleon static properties at finite temperature and density}
In order to gain the chiral soliton solutions and study nucleon static properties at finite temperature and density, the appropriate boundary conditions for the equations of motion in thermal medium need to be defined. As the case in vacuum, the $\sigma$ and $\pi$ are taken as time-independent, classical $c$-number fields, also we require they can only differ from their expectation values in the neighborhood of the quark sources. Therefore, in the absence of the quark source term for large $r$, the equations (\ref{equation5}) and (\ref{equation6}) reduce to the gap equations\cite{Csernai:1995zn}

\begin{eqnarray}
\lambda\left(\sigma^2(r)+\pi^2(r)-\vartheta^2\right)
\sigma(r)-H+g\rho_{s}(r)=0, \label{gap1}\\
\lambda\left(\sigma^2(r)+\pi^2(r)-\vartheta^2\right)
\pi(r)+g \rho_{ps}(r)=0.
\label{gap2}
\end{eqnarray}
with spontaneously broken symmetry, $\sigma (r)=\sigma_v$, and $\pi(r)=0$. After self-consistently solving these gap equations (\ref{gap1}) and (\ref{gap2}) with the constituent quark (antiquark) mass $M_q$ in Eq.(\ref{qmass}), we can get the value of the chiral order parameter $\sigma_v$ for certain $T$ and $\mu$, to which the thermodynamical potential $\Omega(T,\mu)$ has an absolute minimum. Now for the hedgehog baryon, we can safely define the boundary conditions for the coupled nonlinear equations of motion in above as
\begin{eqnarray}
v(0)=0,     \frac{d\sigma(0)}{dr}=0,\pi(0)=0 \label{boundary3}\\
u(\infty)=0,\sigma(\infty)=\sigma_v,\pi(\infty)=0.\label{boundary4}
\end{eqnarray}
and these boundary conditions are indeed satisfied the requirement of finite energy of system, this in turn allow us to define the mass of nucleon (the total energy of the system) in thermal medium properly.

In sequence, according to the boundary conditions Eqs.(\ref{boundary3})(\ref{boundary4}), the set of equations of motion has been solved self-consistently by following the procedure as follows. For a given value of temperature $T$ and density $\mu$, the gap equations (\ref{gap1}) and (\ref{gap2}) can be numerically solved to obtain a constant chiral order parameter $\sigma_v$ which acts as the asymptotic value for $\sigma$ or a mass of the constituent quarks $M_q$. According to this particularly homogeneous solution $\sigma_v$, the above coupled differential equations (\ref{equation1}),(\ref{equation2}),(\ref{equation5}) and (\ref{equation6}) can be solved by the same method used in vacuum but with the normalization condition equation (\ref{norm2}). After that we get the $\sigma$, $\pi$, $u$ and $v$ field configurations as a function of distance radius $r$ for certain $T$ and $\mu$. In the end, by using these solutions, modifications of nucleon properties in thermal medium and the thermodynamics of system could be examined and investigated in detail.

\begin{figure}[thbp]
\epsfxsize=9.0 cm \epsfysize=6.5cm
\epsfbox{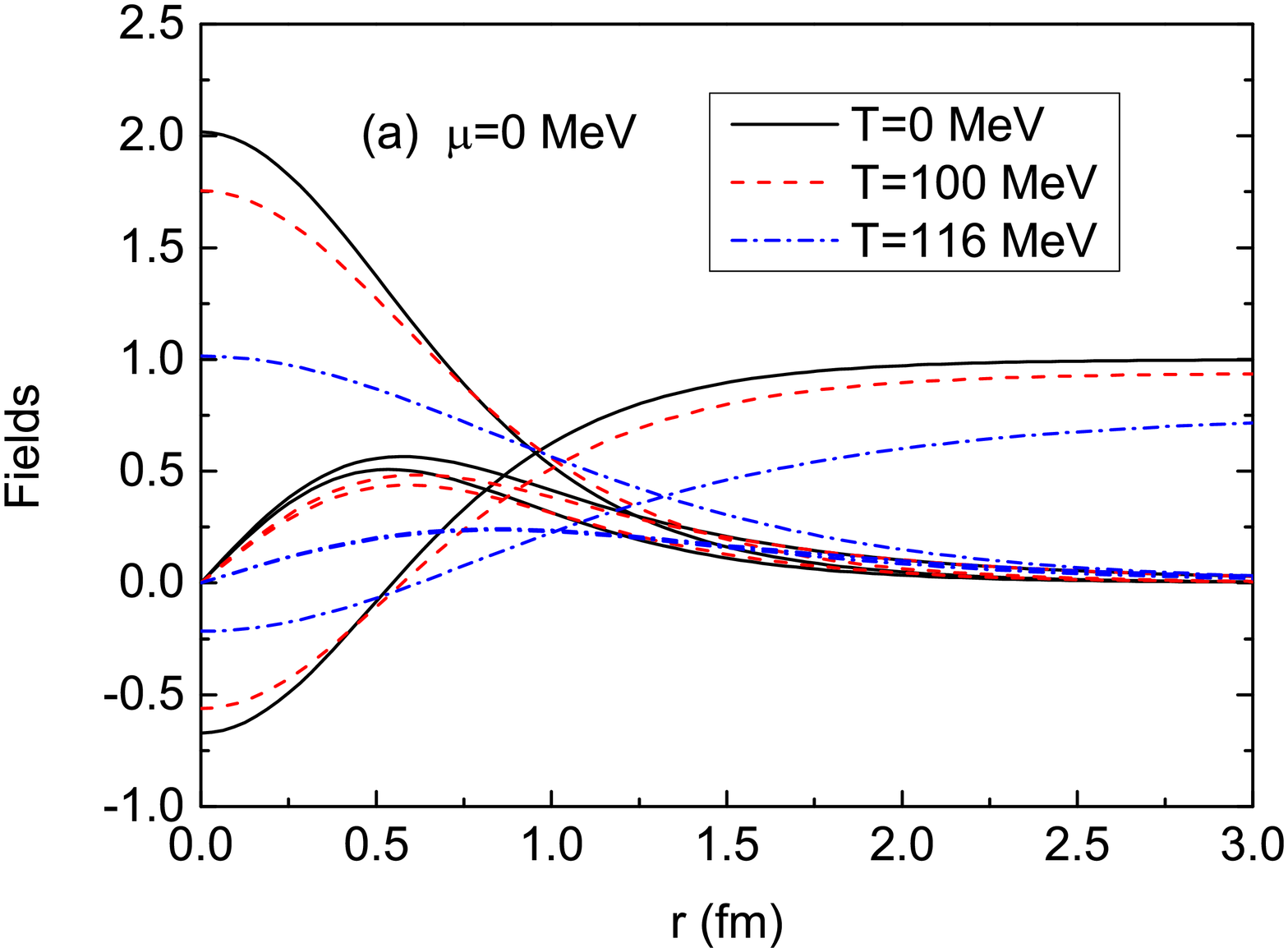}\hspace*{0.1cm} \epsfxsize=9.0 cm
\epsfysize=6.5cm \epsfbox{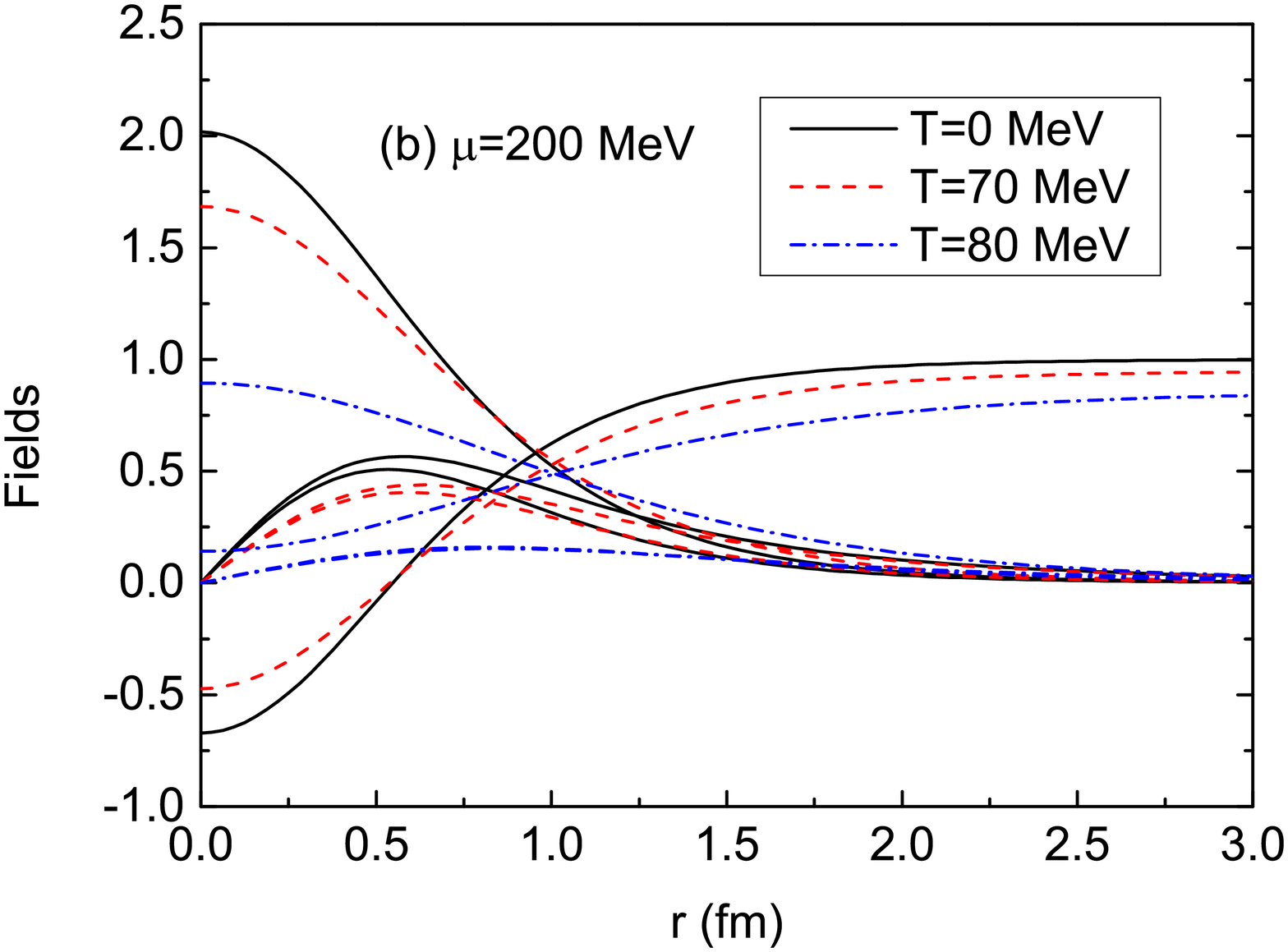}
 \caption{(Color online) (a) The quark fields in relative unit and the $\sigma$, $\pi$ fields scaled with $f_{\pi}$ as function of the radius $r$ at $\mu=0$ MeV, where the solid curves are for $T=0$ MeV, the dashed curves are for $T=100$ MeV and the dash-dotted curves are for $T=116$ MeV. (b) The quark fields in relative unit and the $\sigma$, $\pi$ fields scaled with $f_{\pi}$ as function of the radius $r$ at $\mu=200$ MeV, where the solid curves are for $T=0$ MeV, the dashed curves are for $T=70$ MeV and the dash-dotted curves are for $T=80$ MeV. The parameters are taken as $m_{\sigma}=472$ MeV and $g=4.5$ }
\label{Fig03-04}
\end{figure}

We firstly investigate soliton solutions at finite temperature and density. In Fig.\ref{Fig03-04}, we plot the $u(r)$, $v(r)$, $u(r)$ and $v(r)$ fields at zero and finite chemical potential ($\mu=200$ MeV) for different temperatures. It is shown that all the fields are moving towards to the trivial values with temperature increasing. When $T$ is lager than some critical temperature $T_c$, there only exist the trivial solutions for the coupled equations of motion and solitons are melted away. Furthermore, these trivial solutions indicate the restoration of the chiral symmetry in full space. The lack of solitonic solutions are taken as a signal for the delocalization of the baryonic phase. Sometimes, it is tempting to identify this with deconfinement, but such a conclusion seems to be out of the scope of the model itself, since the linear sigma model only incorporates the chiral symmetry, and according to the present model picture the nucleon is treated as bound state rather than the absolute confinement object.

Base on above analysis, as $T$ is lower than $T_c$, there really exists the baryonic phase, but the stability of this phase should be checked carefully by comparing the total energy of the system in thermal medium with the energy of three free constituent quarks. By subtracting the homogeneous medium contribution\cite{Berger:1996hc}, the total energy of system $E^*$ is given by the sum of the energy of the valence quarks and the kinetic energies of $\sigma$ and $\pi$:
\begin{eqnarray}\label{energy2}
E^*=N \epsilon+4\pi\int r^2 \left[ \frac{1}{2}
\left(\frac{d\sigma}{dr}\right)^2+\frac{1}{2}\left(\frac{d\pi}{dr}\right)^2 \right]dr,
\end{eqnarray}
here, since we take the $\sigma_v$ as a constant for all $r$, the meson interaction energy is discarded. This energy $E^*$ can be considered as an effective mass of nucleon in thermal medium.
\begin{figure}
\includegraphics[scale=0.36]{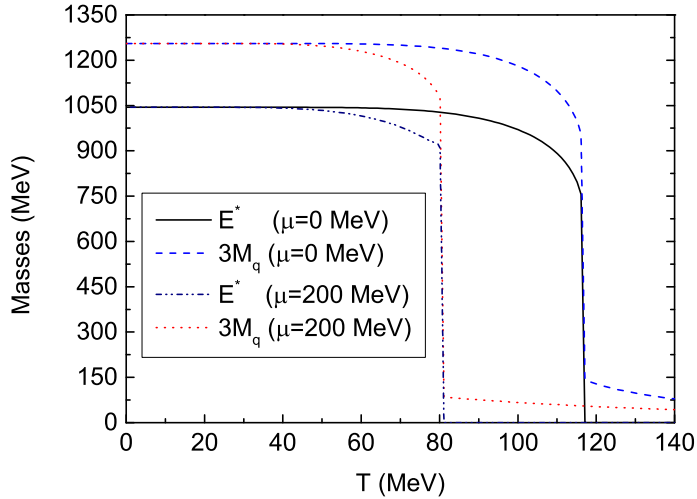}
\caption{\label{Fig05} (Color online) The total energy of system $E^*$ and the energy of $3$ free constituent quark $3M_q$ as function of the temperature $T$ for the parameters taken as $m_{\sigma}=472$ MeV and $g=4.5$. Here one set is for $\mu=0$ MeV, another set is for $\mu=200$ MeV and the meson interaction energy is not included both in vacuum and in thermal medium.}
\end{figure}

In Fig.\ref{Fig05}, the total energy of system $E^*$ is plotted as a function of the temperature for $\mu=0$ MeV and $\mu=200$ MeV, one can see that the total energy $E^*$ monotonically decreases with increasing temperature $T$ from zero to higher value. As the temperature is close to the critical temperature $T_c$, $E^*$ starts to deviate from the ones in vacuum significantly, when $T>T_c$, $E^*$ jumps to zero quickly, which indicates the delocalization phase transition from nucleon matter to quark matter. The energy of three free constituent quark $3M_q$ (or $\sigma_v$) shows the similar behaviors as $E^*$. The critical temperatures are $116$ MeV and $80$ MeV for $\mu=0$ MeV and $\mu=200$ MeV, respectively. These results are very close to previous studies in the Friedberg-Lee model\cite{Reinhardt:1985nq}\cite{Gao:1992zd}\cite{Li:1987wb} or the improved quark mass density-dependent model(IQMDD)\cite{Mao:2006zp}, where a first-order phase transition is predicted and the critical temperature at zero chemical potential is around $100$ MeV. As discussed in Ref.\cite{Scavenius:2000qd}, the critical temperature and the order of phase transition are strongly dependent on the coupling constant $g$. If $g$ gets smaller, then $T_c$ for the chiral phase transition will increase, on some critical value of $g$, the order of chiral phase transition would change from first-order to crossover. The qualitative behaviors of the soliton energy and the constituent quark mass changing with the temperature and density are also found in the Refs.\cite{Christov:1991ry}\cite{Berger:1996hc}\cite{Schleif:1997pi}, where they have used the NJL model instead of the LSM to study chiral solitons\cite{Christov:1995vm}\cite{Weigel:2008zz}. In contrast, our results are different from the work of Abu-Shady and Mansour, in that, by taking the different scheme, they have obtained the nucleon mass monotonically increased with temperature increasing and reached a very large value then slightly decreased, when they took the critical temperature around $161$ MeV. This different result arises from the different treatment on the homogeneous thermal medium contribution, in present discussion, it is subtracted in order to study the stability of the single soliton by comparing with the energy of the three free constituent quarks. However, this contribution should be included if we want to investigate the overall thermodynamics properties of the soliton system, i.e. the internal energy, the entropy and free energy as in Ref.\cite{Schleif:1997pi}.

Comparing the two energies in Fig.\ref{Fig05}, we can show that for $T<T_c$ the nucleon bound sate is stable and $3E_q$ is larger than $E^*$, but the difference decreases with the increasing of temperature, and the two energies start to cross over at the critical temperature $T_c$, after $T_c$, because the chiral symmetry is restored and there is no soliton solution anymore. Hence it is concluded that a chiral phase transition together with a delocalization phase transition from hadron matter to quark matter is going to take place at same critical temperature.

\begin{figure}
\includegraphics[scale=0.36]{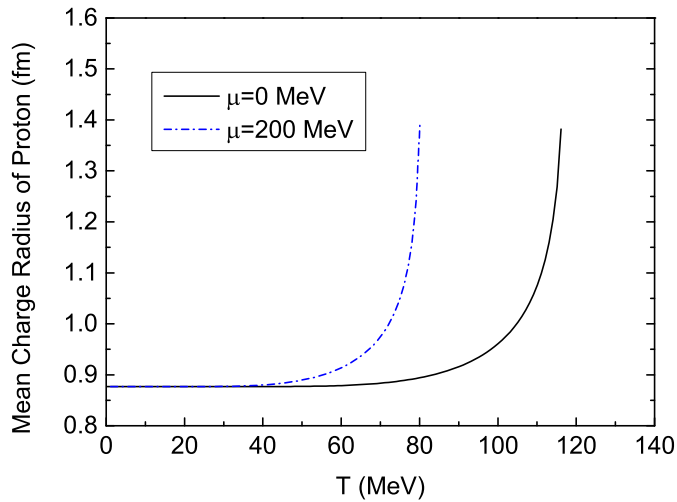}
\caption{\label{Fig06}(Color online) The proton charge `r.m.s' radius of a stable chiral soliton as a function of temperature $T$ at $\mu=0$ MeV and $\mu=200$ MeV for the parameters taken as $m_{\sigma}=472$ MeV and $g=4.5$. The solid curve is for $\mu=0$ MeV while the dash-dotted curve for $\mu=200$ MeV.}
\end{figure}

The proton charge r.m.s radii $R$ of a stable chiral soliton as a function of temperature for $\mu=0$ MeV and $\mu=200$ MeV are illustrated in Fig.\ref{Fig06}, it gives a signal of a swelling of the nucleon when temperature and density increase. In both cases, at low temperature $R$ increases slightly with the increasing of temperature, as $T$ approaching to $T_c$, $R$ will sharply grow and disappear. Another interesting result displayed in Fig.\ref{Fig06} is that the maximal radius $R$ at various densities are almost same when $T$ near $T_c$, this hints that solitons get to overlap each other with the similar expansion rate at $T_c$ for different densities. 

\section{Hadron-quark phase transitions}
In previous section we have obtained the effective nucleon mass at finite temperature and density, for convenience, we set $M_N=E^*$ in the following discussions. In this section we will simply discuss how to calculate the thermodynamical variables of the system from hadron phase to quark phase in the chiral soliton model.

The hadron and quark phase can be distinguished by empirical facts and phenomena at low energy. At low temperature and low baryon density, the hadronic phase exhibits a dynamical breaking of chiral symmetry and the confinement, and the baryon and meson act as the active degrees of freedom here. However, at very high temperature or baryon density, quarks and gluons will be set free to play the dominant roles in QGP. Based on the chiral soliton model, in hadron phase, the free quarks is not the ground state of strongly interacting matter, whereas three valence quarks will form the bound state of the nucleon, then in hadron phase we only have baryons and mesons. On the contrary, when $T>T_c$, the solitons are going to dissolve, then the hadronic phase will eventually evolve to quark phase. In the mean time, when incorporated with the Polyakov-loop, the model can be used to describe thermodynamic properties of QGP satisfactorily\cite{Schaefer:2007pw}\cite{Mao:2009aq}\cite{Schaefer:2009ui}.

\begin{figure}
\includegraphics[scale=0.36]{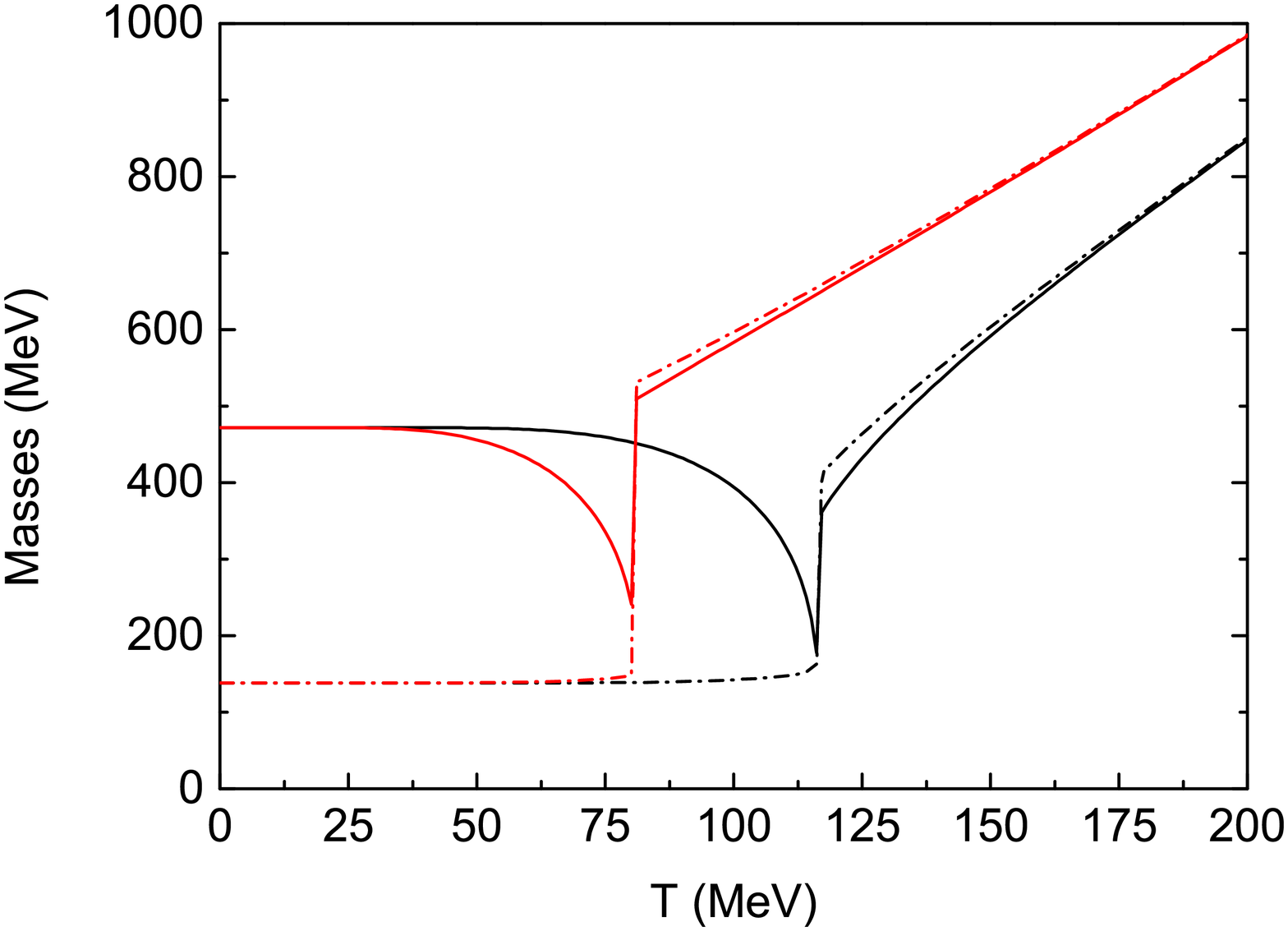}
\caption{\label{Fig07}(Color online) The sigma mass (solid line) and pion mass (dash-dotted line) as a function of temperature for $\mu=0$ MeV (right pair) and for  $\mu=200$ MeV (left pair).}
\end{figure}

Since there is no nucleon in quark phase, all information of the system is contained in the grand canonical potential which is given by $\Omega$ in Eq.(\ref{potential}) with the restoration of chiral symmetry, and the pressure of system is directly given by $P_{QP}=-\Omega_{min}(T,\mu)$. However, in the hadronic phase, the situation becomes a little complicated because the system should be considered as a collection of nuclear matter (solitons) interacted through the self-consistent exchange of $\sigma$, $\omega$, $\rho$ mesons. Moreover, in the spirit of the relativistic mean-field theory (RMF)\cite{Serot:1984ey} or the quark meson coupling model (QMC)\cite{Guichon:1987jp}\cite{Saito:2005rv}, one can investigate the medium modifications of nucleon properties in nuclear matter and finite nuclei extensively\cite{Celenza:1984ew}\cite{Jandel:1983gz}\cite{Wen:2008ui}. This is out of the scope of our topic and left for future study. In following, we merely give an idealistical analysis on this topic by taking the hadronic phase as a noninteracting hadron gas composed of nucleons and $\pi$, $\sigma$ mesons with the effective masses $M_N$, $M_{\pi}$ and $M_{\sigma}$ in thermal medium. Under this scenario, it is straightforward to write down the pressure of the system in terms of nucleons and mesons for the hadronic phase\cite{Yagi:2005yb}\cite{Kapusta:2006pm}
\begin{eqnarray}
P_{HP} &=& \nu_N T \int \frac{d^3\vec{p}}{(2
\pi)^3} \left\{ \mathrm{ln} \left[ 1+e^{-(E_N-\mu_B)/T} \right] +\mathrm{ln} \left[
1+e^{-(E_N+\mu_B)/T}\right]\right\} \nonumber \\&&
-\nu_{\pi} T \int \frac{d^3\vec{p}}{(2
\pi)^3} \left\{ \mathrm{ln} \left[ 1-e^{-E_{\pi}/T}\right]\right\}-\nu_{\sigma} T \int \frac{d^3\vec{p}}{(2
\pi)^3} \left\{ \mathrm{ln} \left[ 1-e^{-E_{\sigma}/T}\right]\right\}-B^*(M_N).
\label{pressure}
\end{eqnarray}
Where $\nu_N=4$ for nucleon, $\nu_{\pi}=3$ for pion and $\nu_{\sigma}=1$ for sigma meson. The last term $B^*(M_N)$ is introduced in order to recover the thermodynamical consistency of the system, since the nucleons are treated as the chiral solitons with a temperature-dependent masses\cite{Gorenstein:1995vm}. The explicit expression of this term can be evaluated by the additional constraint $(\partial P_{HP}/\partial M_N)_T=0$, which gives
\begin{eqnarray}
B^*\left(M_N(T) \right) = B^*\left(M_N(0) \right) -\nu_N \int_0^T dT' \frac{d M_N(T')}{dT'} M_N(T') \int \frac{d^3\vec{p}}{(2
\pi)^3} \frac{1}{E_N'}\left[ \frac{1}{e^{(E_N'-\mu_B)/T'}+1} +\frac{1}{e^{(E_N'+\mu_B)/T'}+1} \right],
\end{eqnarray}
with $E_N'=\sqrt{\vec{p}^2+{M_N(T')}^2}$.

The energies in Eq.(\ref{pressure}) $E_N=\sqrt{\vec{p}^2+{M_N(T)}^2}$, $E_{\pi}=\sqrt{\vec{p}^2+{M_{\pi}(T)}^2}$ and $E_{\sigma}=\sqrt{\vec{p}^2+{M_{\sigma}(T)}^2}$ are corresponding to nucleon, pion and sigma mesons, respectively. $M_N$ is obtained as the energy of soliton, whereas, the $\sigma$ and $\pi$ masses are determined by the curvature of $\Omega(T,\mu)$ in Eq.(\ref{potential}) at the global minimum:
\begin{eqnarray}
M^2_{\sigma}=\frac{\partial^2 \Omega}{\partial \sigma^2},  M^2_{\pi}=\frac{\partial^2 \Omega}{\partial \pi^2}.
\end{eqnarray}
The sigma and pion masses for various $T$ and $\mu$ are shown in Fig.\ref{Fig07}. From this figure, the sigma mass first decreases smoothly and then rebounds and grows again at high temperature. The pion mass does not change much at temperatures below $T_c$ but then increases rapidly, approaching the sigma mass and signaling the restoration of chiral symmetry. At large temperature, the masses grow linearly with the temperature increasing. The sudden jump in Fig.\ref{Fig07} shows there is a first-order phase transition. 

\begin{figure}
\includegraphics[scale=0.36]{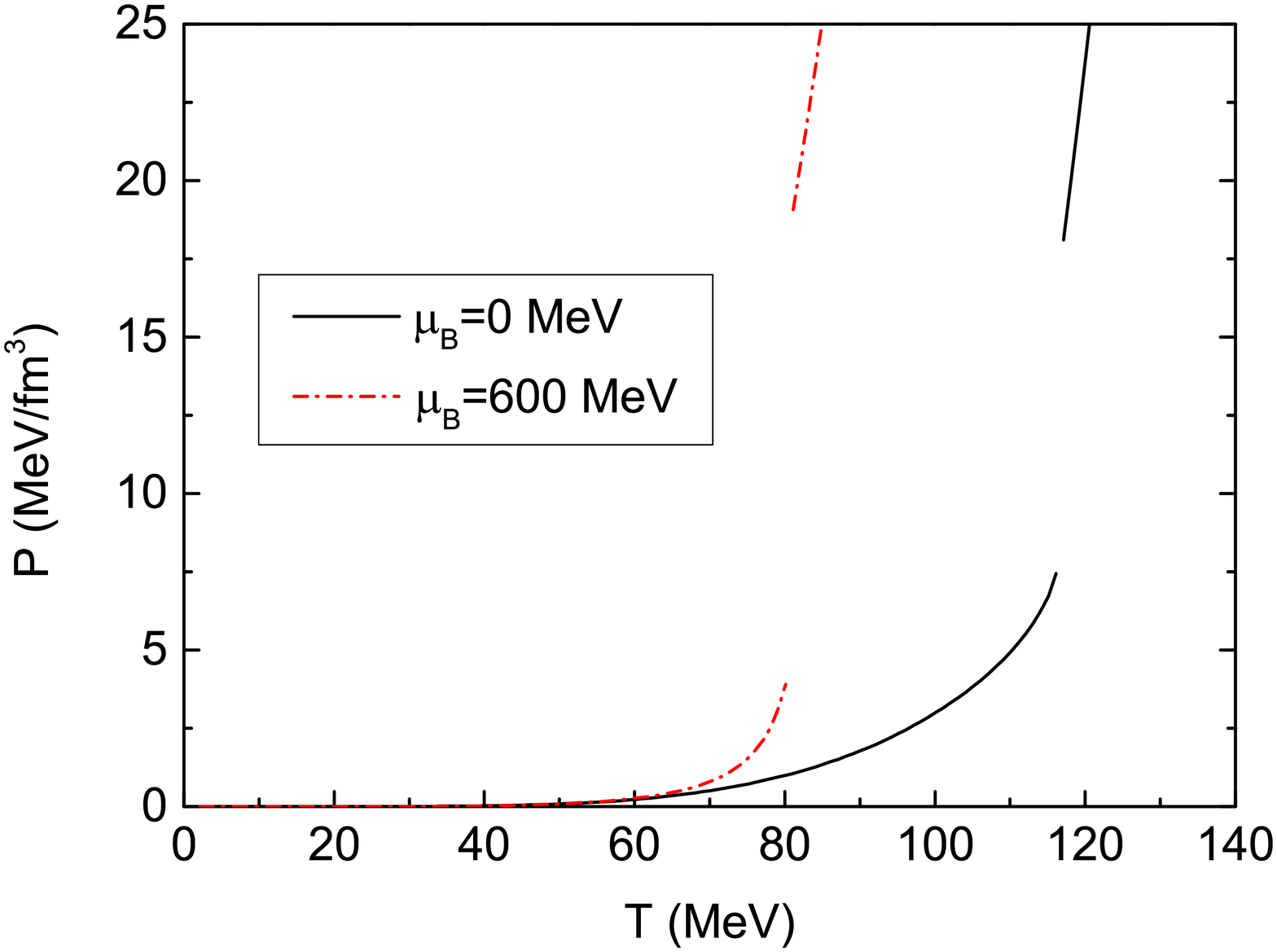}
\caption{\label{Fig08}(Color online) The pressures as a function of temperature for different chemical potential. The solid curve is for $\mu_B=0$ MeV while the dash-dotted curve is for $\mu_B=600$ MeV.}
\end{figure}

With the hadron masses for various temperature and densities, in Fig.\ref{Fig08}, we plot the pressures as a function of the temperature from hadron phase to quark phase at $\mu_B=0$ MeV and $\mu_B=600$. From the figure, all curves show rapidly changed discontinuities at the critical temperature from hadron matter to quark matter. This indicates a first order phase transition for both the chiral phase transition and the delocalization transition, and signals a drastic structural changes in the system. Besides this discovery, we can find that the $P_{HP}$s in the hadronic phase are quite small when compared with those of quark phase, especially, for the case of higher baryon densities. The results can be reflected by the qualitative behaviors of the hadron masses changing with the temperature at various chemical potentials. From Fig.\ref{Fig05}, it is shown that the effective nucleon mass $M_N$ slightly deviates from its vacuum value with temperature increasing, only at the critical temperature $T_c$, $M_N$ experiences a sharp jump to zero. Consequently, the contribution of nucleons to the total pressure in Eq.(\ref{pressure}) is very small in hadron phase as far as the chemical potential is small. For estimate, it only gives $5.2\%$ contribution to $P_{HP}$ when $T$ is around $T_c \approx 116$ MeV and $\mu_B$ is zero. However, with the increasing of the chemical potential $\mu_B$, the contribution of nucleons to the total pressure $P_{HP}$ will become more and more important. The legend can be explained by following two major facts. One is that, from Eq.(\ref{pressure}), the absolute values of the first and second terms in Eq.(\ref{pressure}) will be enhanced if we increase the chemical potential, so do the absolute values of the pressure contributed from nucleons. The other is that, from Fig.\ref{Fig07}, since the minimum masses of sigma meson become larger and larger as $\mu_B$ increasing, the absolute values of the pressure contributed from mesons will decrease accordingly. For illustration, the calculated ratio of the pressure contributed from nucleons to the total one raises up to $29.4\%$ when $\mu_B=600$ as $T$ is around $T_c \approx 80$ MeV, though, in the mean time, the total pressure is less than that of zero chemical potential case. The situation happened in the case of baryons can also be applied for the last term $B^*(M_N)$ in Eq.(\ref{pressure}). For zero and very small chemical potential $\mu_B$, the contribution of the $B^*(M_N)$ to the total pressure $P_{HP}$ only occupied a small part of the total pressure, e.g., as temperature around $T_c$ for $\mu_B=0$, the calculated ratio of the pressure contributed from $B^*(M_N)$ to the total pressure is only about $3.6 \%$ as $B^*(M_N)=0.27 \mathrm{MeV/fm}^3$. But with the increasing of the chemical potential $\mu_B$, the $B^*(M_N)$ would give out a dominant contribution to the total pressure $P_{HP}$. For example, when the temperature is around $T_c\approx 80$ MeV for $\mu_B=600$ MeV, the calculated ratio of the pressure contributed from $B^*(M_N)$ to the total pressure is up to $43.5\%$ as $B^*(M_N)=1.70 \mathrm{MeV/fm}^3$. 

In a whole, the numerically calculated pressures by treating the hadronic phase as a noninteracting hadron gas in hadronic phase are very different from that of the RMF theory at finite temperature and density\cite{Lourenco:2012dx}\cite{Lourenco:2012yv}, where the interactions between hadrons can further reduce the effective masses of nucleons in medium to lower than half of their values in free space. In order to confront the results in RMF theory in hadron phase, the model needs to be improved accordingly by including the interactions of nucleons and mesons with medium-modified coupling constants at hadron level\cite{Saito:2005rv}\cite{Lalazissis:1996rd}, or more practically, considering the possible overlap of nucleons. We believe that these effects will modify the pressures contributed from nucleons and $B^*(M_N)$ dramatically. In the present work, however, we simply take this as a shortcoming of the simple model in the description of nuclear matter.

\section{Summary and discussion}
We have studied the modification of the nucleon properties due to the restoration of the chiral symmetry at finite temperature and density within the linear sigma model with two flavor. The nucleon appears as a chiral soliton in the model, which is embedded in a thermal medium of constituent quarks with self-consistently determined effective mass. The chiral soliton solutions are solved in the mean-field approximation with the restriction to hedgehog configurations. The $T$ and $\mu$ dependent energy of the single soliton is obtained. The only two free parameters $g$ and $m_{\sigma}$ are fixed in order to describe the properties of nucleon in vacuum successfully. The stabilities of the soliton solutions are analysed in thermal medium by comparing the effective mass of nucleon with the energy of three free constituent quarks. Our results show that the chiral phase transition and the delocalization phase transition from nucleon matter to quark matter take place simultaneously. For $T<T_c$, the free constituent quarks are not the ground state of strongly interacting matter, the quarks will develop to form lower-energy bound states carrying the hedgehog configuration. However, as soon as the temperature $T$ crosses over the $T_c$, such bound states can not live anymore, then the system experiences a first-order hadron-quark phase transition to the chirally symmetric phase.

In this work, we predict $T_c$ is about $116$ MeV for $\mu=0$ MeV, but $80$ MeV for $\mu=200$ MeV. It is much lower than the lattice predictions in the range of $150-200$ MeV.  These relatively lower critical temperatures are because of choosing a relative lager coupling constant $g$ which has the functions of making the soliton stable as well as best fitting the proton charge r.m.s radius to $0.877$ fm within the empirical parameter space of $m_{\sigma}$. To overcome this flaw, we can introduce the gluon interaction in the model at quark level, for example, the Polyakov-loop extension of the model\cite{Schaefer:2007pw}. For roughly and qualitatively estimate, assuming the Polyakov loop variable $\Phi(r)$ and its Hermitian conjugate $\bar{\Phi}(r)$ approach to their expectation values as $r\rightarrow \infty$, then we can solve the coupled equations of motion again to get the updated behaviors of the $\sigma_v$ as a function $T$ and $\mu$. By taking the Polyakov loop potential in Ref.\cite{Schaefer:2007pw} with $T_0=190$ MeV, the critical temperature at zero chemical potential now would be about $160$ MeV, it is then quite rational to confront the constraint of the lattice data.

It has turned out that the description of the hadron phase as a non-interacting hadron gas of the nucleons and mesons with medium-modified masses has underestimated the important effects of their interactions at hadron level, and these interactions should be introduced to further reduce the effective masses of nucleons in nuclear matter significantly. Otherwise, such a simple model can not be used to describe the properties of nuclear matter and finite nuclei. Therefore, it is of interest to improve this model to make it suitable for the complete study of the hadron-quark phase transition in the whole region. All these works are in progress.

\begin{acknowledgments}
The authors thank T. Hatsuda and Y. Sakai for fruitful discussions and valuable comments. The work of H.M. and J.J is Supported by Program for Excellent Young Teachers in Hangzhou Normal University, NSFC under No.10904029, 10905014, 11274085 and 11275002.
\end{acknowledgments}

\end{document}